\documentstyle [12pt]{article}
\begin{document}
\begin{center}
{\bf Precision measurements of sodium - sodium and sodium - noble gas 
molecular absorption}
\end{center}

\begin{center}
M. Shurgalin, W.H. Parkinson, K. Yoshino, C. Schoene* and W.P. Lapatovich*
\end{center}

Harvard-Smithsonian Center for Astrophysics, 60 Garden St, MS 14,
Cambridge 
MA 02138 USA

$^{*}$OSRAM SYLVANIA Lighting Research, 71 Cherry Hill Dr, Beverly, MA 
01915 USA

\bigskip

Precision measurements of molecular absorption

\bigskip

PACS numbers: 39.30.+w

07.60.Rd 

33.20.Kf

\bigskip

\begin{center}
Submitted to Measurement Science and Technology, January 2000
\end{center}

\bigskip

\begin{center}
{\bf Abstract.}
\end{center}

\bigskip

Experimental apparatus and measurement technique are described for
precision absorption measurements in sodium - noble gas mixtures. The
absolute absorption coefficient is measured in the wavelength range from
425 nm to 760 nm with $\pm $2\% uncertainty and spectral resolution of
0.02 nm. The precision is achieved by using a specially designed
absorption cell with an accurately defined absorption path length, low
noise CCD detector and double-beam absorption measurement scheme. The
experimental set-up and the cell design details are given. Measurements of
sodium atomic number density with $\pm $5\% uncertainty complement
absorption coefficient measurements and allow derivation of the reduced
absorption coefficients for certain spectral features. The sodium atomic
number density is measured using the anomalous dispersion method. The
accuracy of this method is improved by employing a least-squares fit to
the interference image recorded with CCD detector and the details of this
technique are given. The measurements are aimed at stringent testing of
theoretical calculations and improving the values of molecular parameters
used in calculations.

\bigskip
{\bf Keywords:}  absorption cell, molecular absorption, anomalous 
dispersion method

\bigskip

{\bf 1. Introduction}

\bigskip

Atomic collision processes significantly influence the absorption and
emission of light by atomic vapors at high pressures. As a result the
absorption and emission spectra reveal not only atomic line broadening but
also very broad, essentially molecular features with rich
rotational-vibrational structure and satellite peaks due to formation of
molecules and quasi-molecules. Since pioneering work by Hedges {\it et
al}.  [1], such spectra have been a subject of extensive studies, both
theoretical and experimental, and proved to be a rich source of
information about the interaction potentials, collision dynamics and
transition dipole moments [2-12]. The experimental approaches employed
include absorption measurements [4,5,8,9,12], laser-induced fluorescence
[3,6,9] and thermal emission spectra [7]. While laser-induced fluorescence
and emission spectra provide the shapes and positions of many molecular
bands, the measurements of absorption coefficient spectra also give
absorption coefficients over a large spectral range. Absolute measurements
of the absorption spectra may allow more comprehensive tests of
theoretical calculations. As a result, better differentiation between
different theoretical approaches and improved values for various molecular
parameters and potentials can be obtained.  However, in many cases
absorption spectra are obtained on a relative scale or only the absorption
coefficient or optical depth is measured accurately.  Extraction of
absolute cross-sections (or reduced absorption coefficients)  from
traditional measurements of the optical depth as well as any quantitative
comparisons of absorption spectra with theoretical calculations require
accurate knowledge of the absorption path length and the absorbing species
concentrations.

Most of absorption spectroscopy experiments with hot and corrosive vapors
such as sodium have been performed using heat pipes [8,9,13]. In a heat
pipe the alkali vapor is contained in the hot middle of the furnace
between cold zones where windows are located. Buffer noble gas facilitates
the alkali containment and protects cold windows from alkali deposits. In
this type of absorption cell the vapor density is not uniform at the ends
of the absorption path and the path length is not accurately defined. In
addition, the temperature of the vapor-gas mixture is not uniform and at
higher alkali vapor densities formation of fog at the alkali - buffer gas
interface seriously affects the optical absorption coefficient
measurements [13,14].  Absorption cells have been designed, where heated
windows, placed in the cell hot zone, define the absorption length with
good precision [14-16]. The absorption cell described in [15] is suitable
for hot sodium vapors up to 1000K but it is difficult to make it with a
long absorption path. The cell described in [16] is not suitable for
corrosive vapors and may still have problems with window transmission due
to metal deposits [16]. Schlejen {\it et. al.} [14] designed a cell
specifically for spectroscopy of sodium dimers at high temperatures. Their
cell allowed uniform concentration of absorbers and uniform temperature up
to 1450K with an absorption length defined accurately by hot sapphire
windows. However, the cell design is not suitable for spectroscopy of gas
- vapor mixtures because it was completely sealed and did not easily
enable changing the mixtures.

As well as defining the absorption length accurately, an equally important
aspect is measuring the alkali vapor density. While the noble gas density
can be calculated reasonably well from the measurements of pressure and
temperature using the ideal gas law, it is difficult to establish the
density of alkali atoms. In the majority of reported experiments alkali
concentration was determined from the temperature and published saturated
vapor pressure curves but this approach can introduce significant
uncertainties. For example, in measurements of oscillator strengths or
{\it f}-values significant discrepancies were often obtained between
oscillator strengths measured by methods involving knowledge of the number
density and by methods not requiring it [17]. Even if the vapor pressure
curve is well known for pure saturated vapor, introducing buffer gas or
using unsaturated vapors prohibit accurate knowledge of the vapor density
along the absorption path. To achieve a higher precision in determination
of absolute cross-sections or reduced absorption coefficients it is
necessary to measure the alkali vapor density directly.

In this paper we describe experimental apparatus and technique used for
precision measurements on an absolute scale of molecular absorption
coefficients in sodium vapor + noble gas mixtures. To overcome the
above-mentioned difficulties with definition of absorption length we have
designed a special absorption cell. In our cell, heated sapphire windows,
resistant to hot sodium vapor, are used to define the absorption path. A
high temperature valve, kept at the same temperature as the cell itself,
is utilized to introduce different noble gases. A separate sodium
reservoir, maintained at a lower temperature, is used to control the
sodium vapor pressure independently of the cell temperature. The cell can
be operated at temperatures up to 900K. During the spectral measurements
we measure and monitor the sodium number density at different temperatures
and pressures using the anomalous dispersion or 'hook' method [8,9,18,19].
The 'hook' method allows accurate measurement of {\it nfl} value where
{\it n} is the atomic number density, {\it f} is the atomic line
oscillator strength and {\it l} is the absorption length. If the
absorption length and {\it f}-value for sodium D lines are known, the
sodium number density is accurately obtained. The next section
concentrates on the details of the experiment and the absorption cell
design.

\bigskip

{\bf 2. Experiment}

\bigskip

{\bf 2.1 Experimental set-up}

\bigskip

Fig. 1 shows schematically the experimental set-up that is used for our
absorption measurements. The light source is a 100W halogen lamp powered
by a voltage-stabilized DC power supply. A well-collimated beam of white
light is produced with the help of a condenser lens, an achromat lens, a
pinhole aperture 0.4 mm diameter and another achromat lens of shorter
focal length. The light beam is sent through a Mach-Zender interferometer
and focused on the entrance slit of 3m Czerny-Turner spectrograph
(McPherson model 2163)  with a combination of spherical and cylindrical
lenses. An absorption cell is placed in the test-beam arm of the
Mach-Zender interferometer. Beam blocks are used in both arms to switch
the beams or block them altogether.  The light beam through the reference
arm of the Mach-Zender interferometer is used as a reference for the
absorption in the usual manner of double-beam absorption spectroscopy [9].

The spectra are recorded with a two-dimensional CCD detector (Andor
Technology model V420-0E) placed in the focal plane of the spectrograph.  
This detector has 1024 pixels horizontally and 256 pixels vertically with
pixel size of 26x26 µm. For spectral measurements the detector is used in
the vertical bin mode, that is, as a one-dimensional array detector. The
stigmatic spectrograph has a plane diffraction grating with 2400
grooves/mm and theoretical resolution of $\sim $ 0.005 nm at 500 nm
wavelength. We used 150 µm entrance slit width which gives actual
resolution of 0.02 nm. At least 5 pixels of the array detector are used
over 0.02 nm wavelength interval and as a result smoother spectral data
are obtained from the array detector.

The overall spectral range determined by the diffraction grating and the
detector sensitivity is 425 nm to 760 nm. To record spectra through this
spectral range the diffraction grating is rotated through 160 different
positions by a stepper motor. Backlash is avoided by rotating the grating
in one direction from its calibration position at 425 nm, which in turn is
set by rotating the grating beyond the calibration point. The calibration
point is located by a rotation photosensor placed on the worm screw of the
spectrograph sine-bar mechanism. The photosensor signal is sent to a
programmable stepper motor controller (New England Affiliated Technology
NEAT-310) which drives the stepper motor and allows the grating to be set
at the calibration position automatically.

Each position of the grating permits recording consecutively spectral
intervals ranging from about 3 nm at 430 nm to 1.1 nm at 760 nm, which are
determined by the linear reciprocal dispersion of the spectrograph at a
given wavelength and the overall length of the array detector. All grating
positions are wavelength-calibrated using a large number of different
atomic lines obtained from a number of different hollow cathode spectral
lamps. The wavelength calibration enables identifying the wavelength of
any pixel of the CCD detector within $\pm $ 0.05 nm in the range 425 to
760 nm.

To measure the sodium atomic density using the anomalous dispersion or
'hook' method, both beams through the Mach-Zender interferometer are
unblocked and interfere to produce a spectrally dispersed two-dimensional
interference pattern in the focal plane of the spectrograph. The
Mach-Zender interferometer is adjusted to localize the interference
fringes at infinity.  This insures that the integral sodium number density
along the absorption path is measured. The CCD detector is used in its
normal two-dimensional array mode to record the interference pattern. From
the analysis of the interference pattern recorded in the vicinity of
sodium D lines, sodium number density is derived. The general description
of the 'hook' method is given in [17,18,19] and the details of analyzing
the interference pattern recorded with CCD detector are given in the next
section. A glass plate and a stack of windows, identical to those used in
the absorption cell, are placed in the reference arm of the Mach-Zender
interferometer [17]. These compensating optics remain in the reference
beam during spectral absorption measurements as well and have no effect on
the spectral measurements due to the nature of the dual-beam absorption
technique.

A simple vacuum system consisting of a turbomolecular pump (Sargent Welch
model 3106S) backed by a rotary vane pump (Sargent Welch model 1402) is
used to evacuate the absorption cell. The turbomolecular pump can handle
short bursts of increased pressure and gas flow load and therefore it is
utilized also to pump gases from the cell. A liquid nitrogen trap is
placed in between the cell and the pump to trap sodium vapor. A precision
pressure gauge (Omega Engineering model PGT-45) is used to measure
accurately the pressure of noble gases when filling the cell.

An experiment control and data acquisition computer (Pentium PC) controls
the CCD detector, spectral data acquisition and the spectrograph
diffraction grating via the stepper motor controller connected to the
serial port. The absorption cell temperature is monitored constantly
through a number of thermocouples connected via commercial plug-in data
acquisition board (American Data Acquisition Corporation) and the cell
heaters are controlled via output channels of the data acquisition board
and solid state relays.  Andor Technology CCD software and custom
'in-house' written software are used to perform all these tasks.

\bigskip

{\bf 2.2 Absorption cell}

\bigskip

Fig. 2 shows the schematic diagram of the absorption cell. The cell body
is made of stainless steel (SS) 316 and is approximately 470 mm in length
and 8 mm internal diameter. A vertical extension is welded to the middle
of the cell body, 70 mm in length and 11 mm internal diameter. A sodium
reservoir is located at the end of this extension. It is made of SS 316
with internal diameter 5.5 mm and 70 mm length and it is connected using a
Swagelok fitting which enables disconnection for loading sodium. The
sodium reservoir is heated with a separate heater to introduce sodium
vapor into the cell or it can be cooled with a circulating water cooler to
reduce the alkali number density. Both the heater and cooler are made to
slide on and off the sodium reservoir. A valve is connected to the
vertical extension through which the cell is evacuated and noble gases can
be admitted. This valve is a special bellows-sealed high-temperature valve
(Nupro Company) rated to work at temperatures up to 920K. The valve is
heated to the same temperature or 5 to 10 K higher than the cell itself to
prevent sodium from condensing in the valve.

The major problem one faces when designing an absorption cell with heated
windows is making good vacuum seals for the windows. In case of sodium,
sapphire proved to be material of choice for the windows because of its
excellent resistance to hot sodium [14]. However, it is difficult to make
a reliable sapphire to metal seal that withstands repeated heating cycles
up to 900K. In our design (Fig. 2) the sapphire windows are sealed into
polycrystalline alumina (PCA) tubes with special sealing ceramics (frit)  
used in the construction of commercial high-pressure sodium (HPS) lamps.
The sealing technique used was similar to the one described by Schlejen
{\it et.  al.} [14]. The PCA tubes have 10.2 mm outside diameter and 110
mm length.  They are made from commercial OSRAM SYLVANIA standard HPS
lamps ULX880 by cutting off slightly larger diameter portions. Since PCA
and sapphire have similar thermal expansion coefficients, such window-tube
assembly retains its integrity over a wide range of temperatures. The
tubes are inserted into the heated cell body so that the windows are
located in the hot zone while the tube ends extend to cooler cell ends
where Viton O-rings are used for vacuum seals. Additional external windows
made of fused silica are used with O-ring seals to create an evacuated
transition zone from the heated middle of the cell to the cooler ends.
These silica windows are not heated.

Our cell design allows sodium to condense along the PCA tube towards the
colder zone where O-ring seals are located. To reduce the amount of sodium
condensed there the PCA tubes were carefully selected in outside diameter
tolerance and straightness to match closely the internal diameter of the
SS cell body at room temperature. When the cell is heated the SS expands
more than PCA thus creating some space for sodium to condense. Once the
sodium build-up reaches the hot zone, no more sodium is lost into the void
along the PCA tubes.

The windowed ends of the PCA tubes rest against the stepped profiles
inside the SS cell body as shown in Fig. 2. These stepped profiles
determine the positions of the heated windows and thus the absorption
length. To ensure that the PCA tubes are always firmly pressed against
these stepped profiles regardless of the thermal expansion differences
between PCA and SS, compression springs are used to push the PCA tubes via
stacks of spacers, made of SS 316, and the external windows. Cap nuts
complete the assembly of the windows, PCA tubes, O-ring seals and spacers
as Fig. 2 illustrates.  These caps allow easy removal of all windows for
cleaning if needed as well as adjustment of the spring compression. The
compression springs are chosen to produce about 12 N force, equivalent to
about 1.5 atm on the surface area of the heated windows.

The absorption length at room temperature is measured using a special tool
made of two rods about 4 mm in diameter inserted into a tube of 6 mm
outside diameter. One rod is permanently fixed while the other can slide
in and out with friction, thus allowing change in the overall length of
the tool. The ends of the tool are rounded and polished. With one sapphire
window completely in place at one end of the cell, the tool is inserted
into the cell and the second PCA tube with sapphire window is put in
place. The tool adjusts its length precisely to the distance between two
sapphire windows.  Then the PCA tube and the tool are carefully taken out
and the length of the tool is measured with a micrometer. In our cell the
absorption length at room temperature was measured 190.03 $\pm $ 0.025 mm.
The absorption length at operating temperature is calculated from the
temperature of the cell and the thermal expansion coefficient for SS of 18
$\pm $ 2.2 x 10$^{-6}$ K$^{-1}$ [20]. Since the change in the absorption
path length due to thermal expansion is a small percentage of the overall
length, large uncertainties in the thermal expansion calculation do not
lead to a large uncertainty in the resulting absorption length at a given
temperature.

The whole absorption cell including the valve is heated by sets of heaters
made of Nickel-Chromium wire. Separate sets of heaters are used to heat
the cell and the valve. Each heater set consists of two separate heaters.
One is switched on constantly while the other one is used in on-off mode,
controlled from the experiment control computer, to maintain average cell
and valve temperatures constant. Six type K thermocouples are used to
measure the temperatures at different points. Three thermocouples are
placed in contact with the main cell body, one of them in the middle and
the other two at the locations of heated windows. Another thermocouple
measures the valve temperature. Two thermocouples are used to measure the
temperatures at the bottom and at the top of the sodium reservoir. The
heaters are isolated from each other and the SS parts of the cell by
embedding them into insulation made of moldable Alumina-Silica blankets
(Zircar Products). All heated parts are wrapped into thermal insulation
made of Alumina blankets (Zircar Products). The positions of the heaters
are chosen as shown schematically in Fig. 2. The middle part of the cell
of $\sim $60 mm length does not have heaters but is nevertheless heated
sufficiently by thermal conductance. Also the thermal insulation is
adjusted in such a way that the temperature, measured with thermocouple in
the middle of the cell, is 5 to 10 K lower than the temperature at the
points where the heated windows are located. Heating sapphire windows to a
slightly higher temperature ensures they remain free from any deposits
during the operation of the cell.

The cell body thermocouple and the valve thermocouple readings give an
average temperature of the cell body. The gas mixture temperature is
assumed to be equal to the average temperature of the cell body. Since the
thermocouples are located between the SS cell body and the heaters, they
may give readings of slightly higher temperature than the actual
temperature of the cell and the gas inside it. Given this fact and the
temperature reading differences between the thermocouples, the uncertainty
in the gas mixture temperature is estimated to be + 10 K - 50 K.

\bigskip

{\bf 2.3 Measurement technique}

\bigskip

Absorption spectroscopy is based on Beer's law describing absorption of
light in homogeneous absorbing media

\begin{equation}
I_{1} \left( {\lambda}  \right) = I_{0} \left( {\lambda}  \right)exp\left(
{ 
- k\left( {\lambda}  \right)l} \right)
\end{equation}

\noindent where {\it I}$_{1}$ is the transmitted intensity of light, {\it
I}$_{0}$ is the incident intensity of light, {\it k} is absorption
coefficient and {\it l} is the absorption length. In real experimental
measurements the transmission through optics, absorption cell windows and
spectrograph, detector sensitivity and light source spectral
characteristics all have to be taken into account. In the dual beam
arrangement for absorption spectroscopy the test $S_{t} \left( {\lambda}
\right)$ and reference $S_{r} \left( {\lambda} \right)$ beam spectra,
obtained from the detector, are

\begin{equation}
\label{eq1}
S_t (\lambda ) = k_t^0 (\lambda )I_0 (\lambda )\exp \left( { - \tau _t
(\lambda ) - k(\lambda )l} \right)
\end{equation}

\begin{equation}
\label{eq2}
S_{r} \left( {\lambda}  \right) = k_{r}^{0} \left( {\lambda}  \right)I_{0} 
\left( {\lambda}  \right)exp\left( { - \tau _{r} \left( {\lambda}
\right)} 
\right)
\end{equation}

\noindent where {\it I}$_{0}$ is the intensity of the light source,
$k\left( {\lambda } \right)$is the absorption coefficient to be measured,
{\it l} is the absorption length and $k_{t}^{0} $, $\tau _{t} $ and
$k_{r}^{0} $, $\tau _{r} $ are the coefficients that take into account the
detector efficiency, absorption of all optics elements such as windows and
lenses and spectrograph transmission. To eliminate all unknown parameters
represented in (\ref{eq1}) and (\ref{eq2}) by $k_{t}^{0} $, $\tau _{t} $,
$k_{r}^{0} $ and $\tau _{r} $, we measure first the reference spectra (the
spectra obtained without sodium in the absorption path and thus without
atomic and molecular absorption we are interested in). Sodium
concentration in the absorption path is reduced to less than 10$^{14}$
cm$^{-3}$ by cooling the sodium reservoir down to between +5 to + 10° C
using the circulating water cooler around it. At densities below 10$^{14}$
cm$^{-3}$ the molecular absorption of both sodium-sodium and sodium-noble
gas is negligible and $k\left( {\lambda} \right) = 0$. Both test and
reference beam spectra are taken at each grating position and their ratio

\begin{equation}
\label{eq3}
\frac{{S_{t}^{0} \left( {\lambda}  \right)}}{{S_{r}^{0} \left( {\lambda}  
\right)}} = \frac{{k_{t}^{0}} }{{k_{r}^{0}} }exp\left( {\tau _{r} \left( 
{\lambda}  \right) - \tau _{t} \left( {\lambda}  \right)} \right)
\end{equation}

\noindent is calculated. Thus obtained reference spectra contain
information about all unknown parameters. To reduce statistical error a
number of measurements are performed and the average is calculated.

To measure the absorption spectra of sodium-sodium and sodium-noble gas
molecules, the sodium vapor is introduced in the absorption path by
heating the sodium reservoir. Both test and reference beam spectra are
taken at each diffraction grating position and their ratio is calculated:

\begin{equation}
\label{eq4}
\frac{{S_{t}^{Na} \left( {\lambda}  \right)}}{{S_{r}^{Na} \left( {\lambda}  
\right)}} = \frac{{k_{t}^{0}} }{{k_{r}^{0}} }exp\left( {\tau _{r} \left( 
{\lambda}  \right) - \tau _{t} \left( {\lambda}  \right) - k\left(
{\lambda 
} \right)l} \right)
\end{equation}

Once again to reduce statistical error a number of measurements are
performed and averaged. From (\ref{eq3}) and (\ref{eq4}) it follows that
the absorption coefficient {\it k}($\lambda $) is obtained from
measurements of absorption and reference spectra with all unknown
parameters eliminated:

\begin{equation}
k\left( {\lambda}  \right) = - \frac{{1}}{{l}}ln\left( 
{{\raise0.7ex\hbox{${\frac{{S_{t}^{Na}} }{{S_{r}^{Na}} }}$} 
\!\mathord{\left/ {\vphantom {{\frac{{S_{t}^{Na}} }{{S_{r}^{Na}} }} 
{\frac{{S_{t}^{0}} }{{S_{r}^{0} 
}}}}}\right.\kern-\nulldelimiterspace}\!\lower0.7ex\hbox{${\frac{{S_{t}^{0} 
}}{{S_{r}^{0}} }}$}}} \right)
\end{equation}

Using the procedure described above we are able to measure the absolute
absorption coefficient with as small as $\pm $1 \% statistical error in
the range 425 - 760 nm with spectral resolution $\sim $0.02 nm.

Derivation of the reduced absorption coefficient for sodium-sodium and
sodium-noble gas quasi-molecules requires accurate knowledge of atomic
number densities. The atomic density for noble gas is calculated from
pressure and temperature using the ideal gas relationship and the sodium
density is measured by the 'hook' method [17-19]. Fig. 3 shows the 'hook'
interference pattern recorded with CCD detector in the focal plane of the
spectrograph. The analysis of this pattern and extraction of the sodium
atomic number density is performed by a least-squares fit of the
interference fringe model to the recorded pattern using software
specifically written for this purpose. The following equation can be used
to describe the positions $ y_{k} $ of interference fringes of maximum
intensity in the focal plane of the spectrograph [19]:

 $ y_{k} = a\left( {k\lambda - \left( {\frac{{A_{1}} }{{\lambda _{1} -
\lambda 
}} - \frac{{A_{2}} }{{\lambda _{2} - \lambda} }} \right)N + \Delta nl} 
\right) \quad , $

$\quad
A_{1} = \frac{{r_{0} g_{1} f_{1} l\lambda _{1}^{3}} }{{4\pi} }$ and $A_{2}
= 
\frac{{r_{0} g_{2} f_{2} l\lambda _{2}^{3}} }{{4\pi} }$

\noindent where $r_{0} $ is the classical electron radius, $g_{1} , f_{1}
, \lambda _{1} $ are respectively the{\it g}-factor, the oscillator
strength and the wavelength of sodium D1 line, $g_{2} ,f_{2} ,\lambda _{2}
$ are respectively the {\it g}-factor, the oscillator strength and the
wavelength of sodium D2 line, {\it l} is the absorption path length,
$\Delta n$ is the coefficient accounting for optical path length
difference between test and reference beams of the Mach-Zender
interferometer, {\it a} is the scaling factor accounting for imaging
properties of the optical set-up, {\it k} is the fringe order and {\it N}
is the sodium number density.$_{} $ The above equation is valid at
wavelengths separated from the atomic line core by more than the FWHM of
the broadened line, $\lambda _{} - \lambda _{i} > > \Delta \lambda $ [17].
Our model calculations of the 'hook' interference pattern, which included
the atomic line broadening, showed that the error introduced by neglecting
the atomic line broadening in the above equation, is negligible when
atomic number density of sodium is above 5x10$^{14}$ cm$^{-3}$ and noble
gas pressure is below 500 Torr. These conditions are always met in our
measurements. After some algebraic manipulations the following fringe
model equation is obtained which gives positions $y_{i} $ of a number of
fringes in terms of 2D CCD detector coordinates and fit parameters:

\begin{equation}
y_i  = a_3  + a_2 \lambda  + a_1 i\lambda  + a_4 \left( {\frac{{A_1
}}{{\lambda _1  - \lambda }} + \frac{{A_2 }}{{\lambda _2  - \lambda }}}
\right)
\end{equation}

\noindent 

where $y_{i} $ is the vertical fringe coordinate at a given wavelength
$\lambda $, {\it i} = 0,1,2,3,4,5 denotes adjacent fringes seen by the
detector and $a_{1} $, $a_{2} $, $a_{3} $ and $a_{4} $ are the fit
parameters. The sodium number density is calculated from fit parameters
$a_{1} $ and $a_{4} $:

\begin{equation}
N = \frac{{a_{4}} }{{a_{1}} }
\end{equation}

From the recorded interference pattern three to five interference fringes,
defined at maximum intensity, are extracted at each side of the sodium
doublet and the ({\it y, $\lambda $}) coordinates for each fringe are
calculated from the CCD pixel coordinates and wavelength calibration to
provide the data set for the least-squares fit. Fig. 4a shows the
interference fringes obtained from the image presented in Fig. 3 and Fig
4b shows the fitted model curves. Since a large number of points are used
to locate the fringe positions, higher accuracy can be achieved in
comparison with traditional methods of extracting atomic number density
from measuring the location of the 'hook' maxima of a single fringe [17].
The main limitation on the accuracy of this new technique is imposed by
wavelength calibration, especially at lower atomic densities. Using the
system described above, the sodium density is routinely measured with $\pm
$ 2 \% uncertainty, given the wavelength uncertainty $\pm $0.03 nm in the
vicinity of the sodium D-lines and the uncertainty in the interference
fringe position of $\pm $5 pixels. During consecutive spectral
measurements used in calculating the resultant average spectra, the sodium
number density was measured at the beginning of each measurement and was
found to remain constant within $\pm $ 4\% to $\pm $5\%.

\bigskip

{\bf 3. Measurement results}

\bigskip

Fig. 5 presents a spectrum of the absolute absorption coefficient of a
sodium - xenon mixture measured at $900K^{+10K}_{-50K} $ cell temperature.
The Xe pressure is 400 $\pm $ 0.5 Torr, which gives xenon density
4.29x10$^{18}$ cm$^{-3}$ at 900 K temperature. Sodium density is measured
as 2.05 $\pm $ 0.06 x 10$^{16}$ cm$^{-3}$. The absorption coefficient in
the 425 nm to 760 nm range consists of contributions from the broadened
sodium atomic lines around 589 nm, the sodium - noble gas and the sodium -
sodium molecular spectra. From 460 nm to about 550 nm a blue wing of the
sodium dimer absorption is apparent [5]. At 560 nm there appears a
sodium-xenon blue wing satellite feature [6] and towards the longer
wavelength of the significantly broadened sodium D-lines there are red
wings of the sodium dimer [5] and the sodium-xenon molecules [6]. Fig 6
shows a spectrum of the absolute absorption coefficient of a sodium -
argon mixture measured at $900K^{+10K}_{-50K} $ cell temperature. The Ar
pressure is 401 $\pm $ 0.5 Torr, which gives argon density 4.3x10$^{18}$
cm$^{-3}$ at 900 K temperature. Sodium density is measured as 1.00 $\pm $
0.04 x 10$^{16}$ cm$^{-3}$. This spectrum is similar to the sodium - xenon
spectrum shown in Fig. 5 except that the sodium - argon blue wing
satellite is located at a slightly shorter wavelength of 554.5 nm and the
sodium - argon red wing extends further from the sodium atomic line core.
The magnitude of the absorption coefficient is lower in proportion to the
lower sodium density.

Fig. 7 illustrates rotational-vibrational features of sodium dimer
absorption at a 0.02 nm resolution in the vicinity of the 520 nm band. The
features are a complicated superposition of many rotational-vibrational
bands of the sodium dimer and identification of these bands has not yet
been attempted. The statistical uncertainty in the absorption coefficient
magnitude is indicated. This uncertainty includes both detector
statistical errors and the uncertainty in the absorption path length. At
any wavelength in the 425 nm to 760 nm range the uncertainty in the
absorption coefficient does not exceed $\pm $2\% where absorption
coefficient values are larger than 0.008 cm$^{-1}$. The measured spectra
can serve for stringent quantitative tests of theoretical calculations
[21]. Preliminary comparisons showed good overall agreement between the
measurements and theoretical calculations at a temperature of 870 K, which
is within our experimental temperature uncertainty [22]. Full details of
the calculations and comparisons with experiment will be presented in the
forthcoming publication [23].

A reduced absorption coefficient is calculated for the blue wing of sodium
dimer absorption, which is well separated from the rest of the spectrum,
using the measured absorption coefficient and sodium atomic number density
and it is presented in Fig. 8. Apart from the broad and strong molecular
absorption arising mostly from transitions from bound to bound states
between $X\quad {}^{1}\Sigma _{g}^{+} $ and $B\quad {}^{1}\Pi _{u}^{} $
molecular singlet states of the sodium dimer, there are also features from
the triplet transitions $2\quad {}^{3}\Pi _{g}^{} \leftarrow a\quad
{}^{3}\Sigma _{u}^{+} $ and $c{}^{3}\Pi _{g} \leftarrow a\quad
{}^{3}\Sigma _{u}^{+} $ [5,24].

Since the sodium-xenon molecular absorption bands are very close to the
sodium D-lines, it is difficult to separate them completely from atomic
lines and to derive the reduced absorption coefficient. Fig.9 presents the
absolute absorption coefficient in the vicinity of sodium-xenon blue
satellite features at different xenon densities and sodium density of
7.7x10$^{15}$ cm$^{-3}$ and at 900 K temperature. There are two satellite
features present at 560 nm and 564 nm. The positions and relative
magnitude of these features can provide some insights into potentials of
the sodium-xenon molecule as well as transition dipole moments [21,23].

\bigskip
{\bf 4. Conclusion}

\bigskip

Details of precision absorption measurements in sodium - noble gas
mixtures at high spectral resolution have been presented. To perform more
stringent tests of theoretical calculations and molecular parameters used
in the calculations the goal was to obtain the absorption coefficient
spectra on an absolute scale with better than $\pm $ 2\% uncertainty at
near UV and visible wavelengths. To achieve such precision an absorption
cell for containment of sodium vapor with accurately defined absorption
path was constructed. The measurements were performed using double-beam
absorption measurement scheme eliminating all unknown parameters such as
detector sensitivity and optics transmission. A low noise CCD detector was
used to record the spectra and a number of separate measurements were
averaged to reduce statistical error. To measure accurately the alkali
number density the anomalous dispersion or 'hook' method was employed. The
accuracy of the 'hook' method was improved by means of least-squares fit
to the interference fringes image recorded using 2D CCD detector in the
focal plane of the spectrograph. The measurements obtained with the
apparatus and technique described extend the available data on the sodium
- sodium and sodium - rare gas absorption to different temperatures and
higher precision and spectral resolution.

\bigskip
{\bf 5. Acknowledgements}

\bigskip

This work is supported in part by National Science Foundation under grant
No PHY97-24713. The authors would like to acknowledge useful discussions
with J.F. Babb, H. Adler and G. Lister and generous equipment and
materials support from OSRAM SYLVANIA.

\bigskip

{\bf References}

\bigskip

[1] R.E. Hedges, D.L. Drummond and A. Gallagher 1972 {\it Phys. Rev. A}
{\bf 6}, 1519

[2] D.L. Drummond and A. Gallagher 1974 {\it J. Chem. Phys.} {\bf 60},
3246

[3] W. Demtröder and M. Stock 1975 {\it J. Mol. Spectr.} {\bf 55}, 476 

[4] J.F. Kielkopf, and N.F. Allard 1980 {\it J. Phys. B} {\bf 13}, 709

[5] J. Schlejen, C.J. Jalink, J. Korving, J.P. Woerdman and W. Müller 1987
{\it J. Phys. B} {\bf 20}, 2691

[6] K.J. Nieuwesteeg, Tj. Hollander and C.Th. J. Alkemade 1987 {\it J.
Phys.  B} {\bf 20}, 515

[7] J. Schlejen, J.P. Woerdman and G. Pichler 1988 {\it J. Mol. Spectr.}
{\bf 128}, 1

[8] K. Ueda, H. Sotome and Y. Sato 1990 {\it J. Chem. Phys.} {\bf 94},
1903

[9] K. Ueda, O. Sonobe, H. Chiba and Y. Sato 1991 {\it J. Chem. Phys.}
{\bf 95}, 8083 

[10] D. Gruber, U. Domiaty, X. Li, L. Windholz, M.
Gleichmann and B. A. Heß 1994 {\it J. Chem. Phys.} {\bf 102}, 5174

[11] J. Szudy and W.E. Baylis 1996 {\it Phys. Rep.} {\bf 266}, 127

[12] P.S. Erdman, K.M. Sando, W.C. Stwally, C.W. Larson, M.E. Fajardo 1996
{\it Chem. Phys, Lett.} {\bf 252}, 248

[13] A. Vasilakis, N.D. Bhaskar and W Happer 1980 {\it J. Chem. Phys.}
{\bf 73}, 1490

[14] J. Schlejen, J. Post, J. Korving and J.P. Woerdman 1987 {\it Rev.
Sci.  Instrum.} {\bf 58}, 768

[15] A.G. Zajonc 1980 {\it Rev. Sci. Instrum.} {\bf 51}, 1682

[16] Y. Tamir and R. Shuker 1992 {\it Rev. Sci. Instrum.} {\bf 63}, 1834

[17] W.H. Parkinson 1987 {\it Spectroscopy of Astrophysical Plasmas
(Cambridge University Press)}

[18] D. Roschestwensky 1912 {\it Ann. Physik,}{\bf 39}, 307

[19] M.C.E. Huber and R.J. Sandeman 1986 {\it Rep. Prog. Phys.} {\bf 49}
397 

[20] American Institute of Physics Handbook 1972 4-138 {\it (McGraw
Hill Book Company)}

[21] H-K. Chung, M. Shurgalin and J.F. Babb 1999 {\it 52}$^{nd}${\it GEC,
Bull. APS}, {\bf 44}, 31

[22] H-K. Chung and J.F. Babb, {\it private communication}

[23] H-K. Chung, K. Kirby, J.F. Babb and M. Shurgalin, 2000, {\it in
preparation}

[24] D.Veza, J. Rukavina, M. Movre, V. Vujnovic and G. Pilcher 1980 {\it
Optics Comm.} {\bf 34} 77

\bigskip

{\bf Figure 1.} Schematic diagram of the experimental set-up.

\bigskip

{\bf Figure 2.} Schematic diagram of the absorption cell.

\bigskip

{\bf Figure 3.} Image of a 'hook' pattern obtained with a two-dimensional
CCD detector.

\bigskip

{\bf Figure 4.} Interference fringes extracted from 'hook' pattern image
(a)  and fitted model curves (b).

\bigskip

{\bf Figure 5.} Absorption coefficient of sodium - xenon mixture at 900 K.

\bigskip

{\bf Figure 6.} Absorption coefficient of sodium - argon mixture at 900 K.

\bigskip

{\bf Figure 7.} Rotational-vibrational features of sodium dimer absorption
spectra at 0.02 nm resolution.

\bigskip

{\bf Figure 8.} Reduced absorption coefficient for the blue wing of sodium
dimer molecular absorption.

\bigskip

{\bf Figure 9.} Blue wing of sodium-xenon molecular absorption.

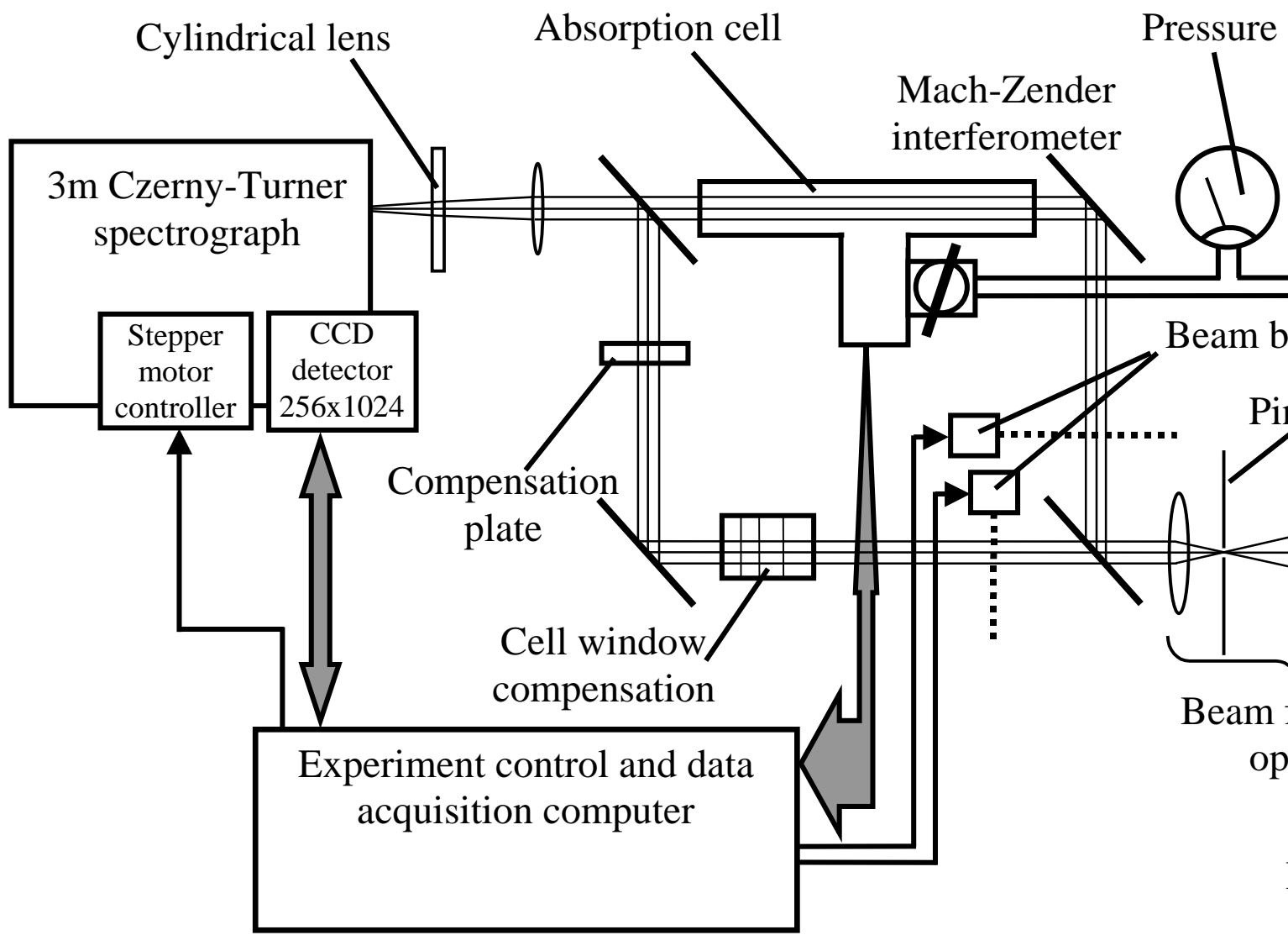
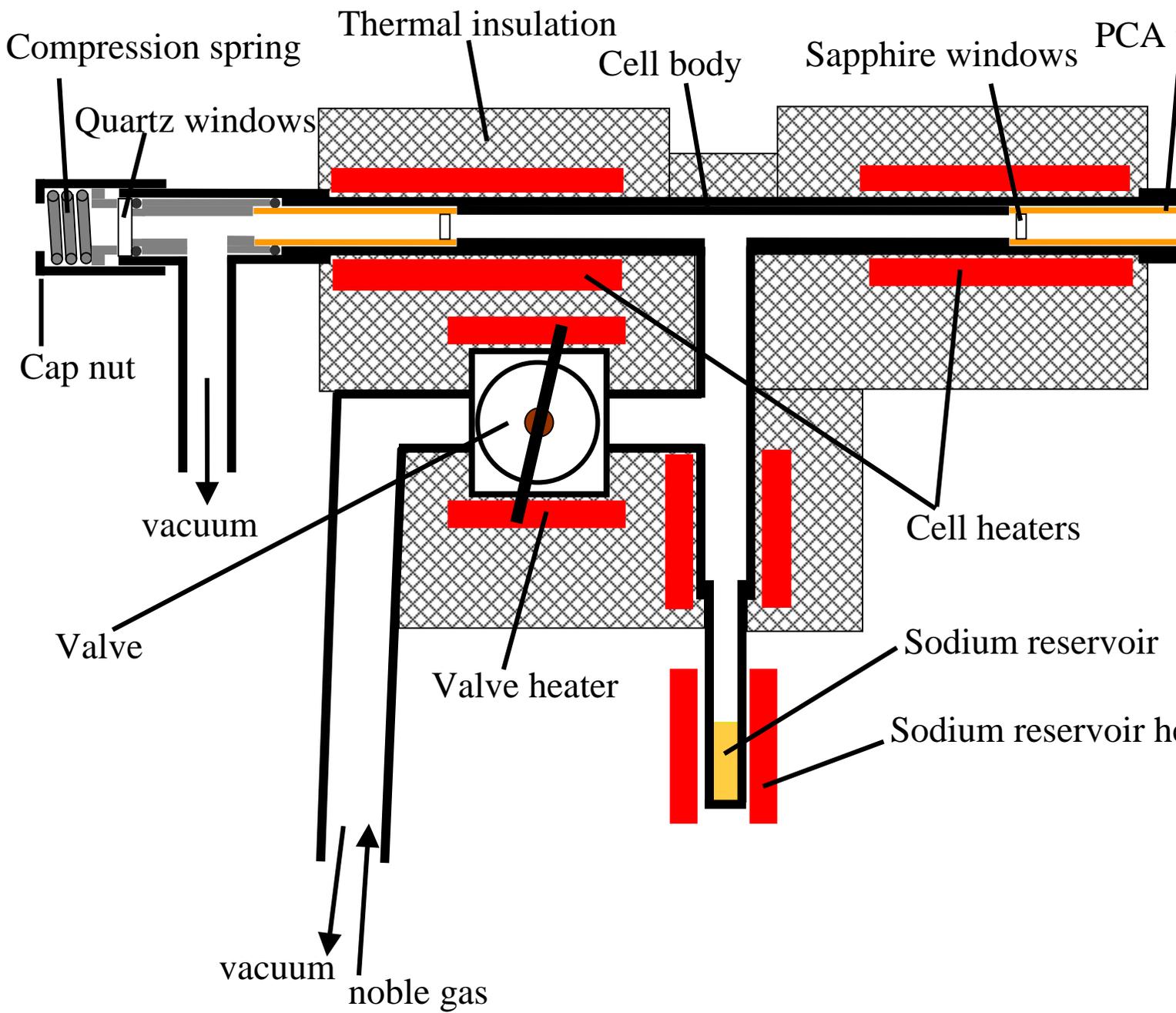
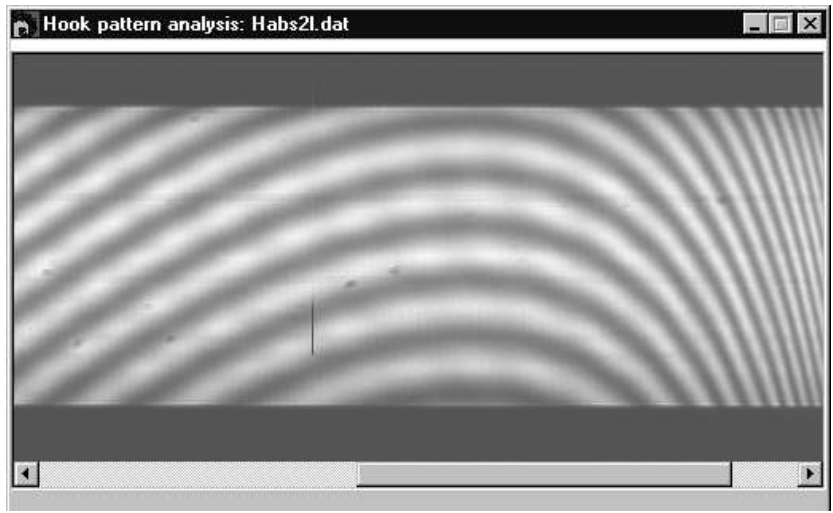
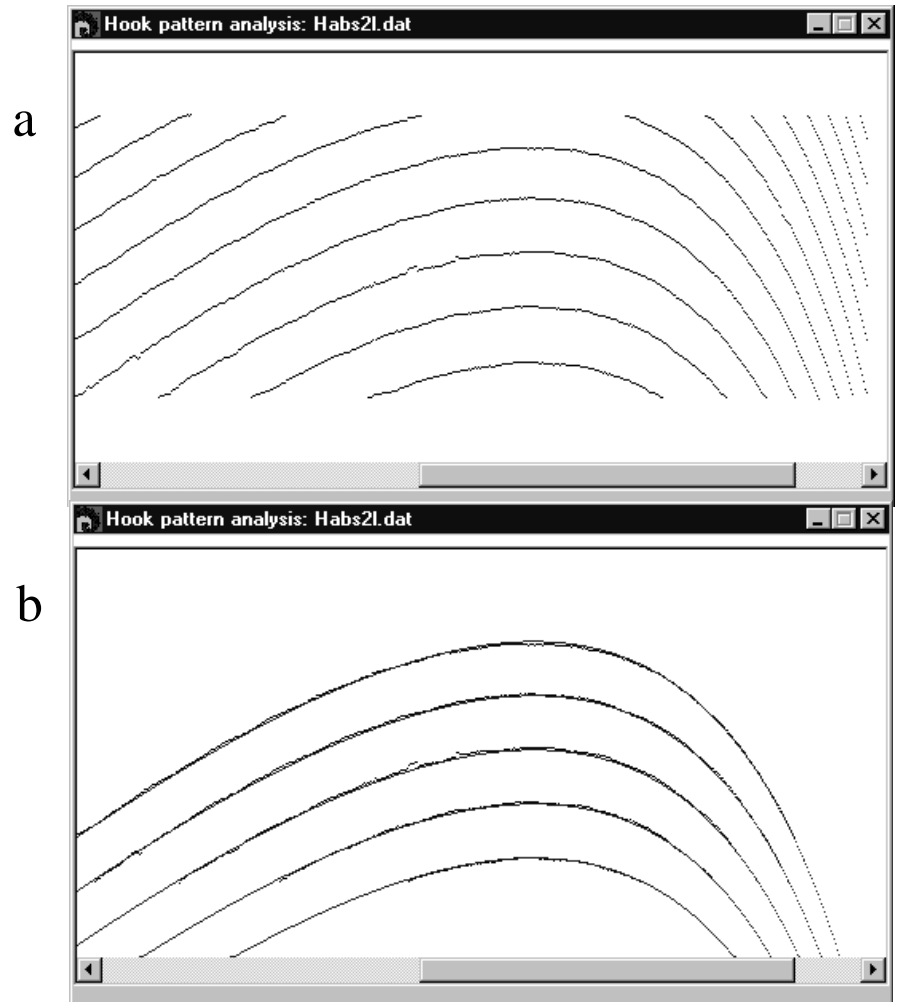
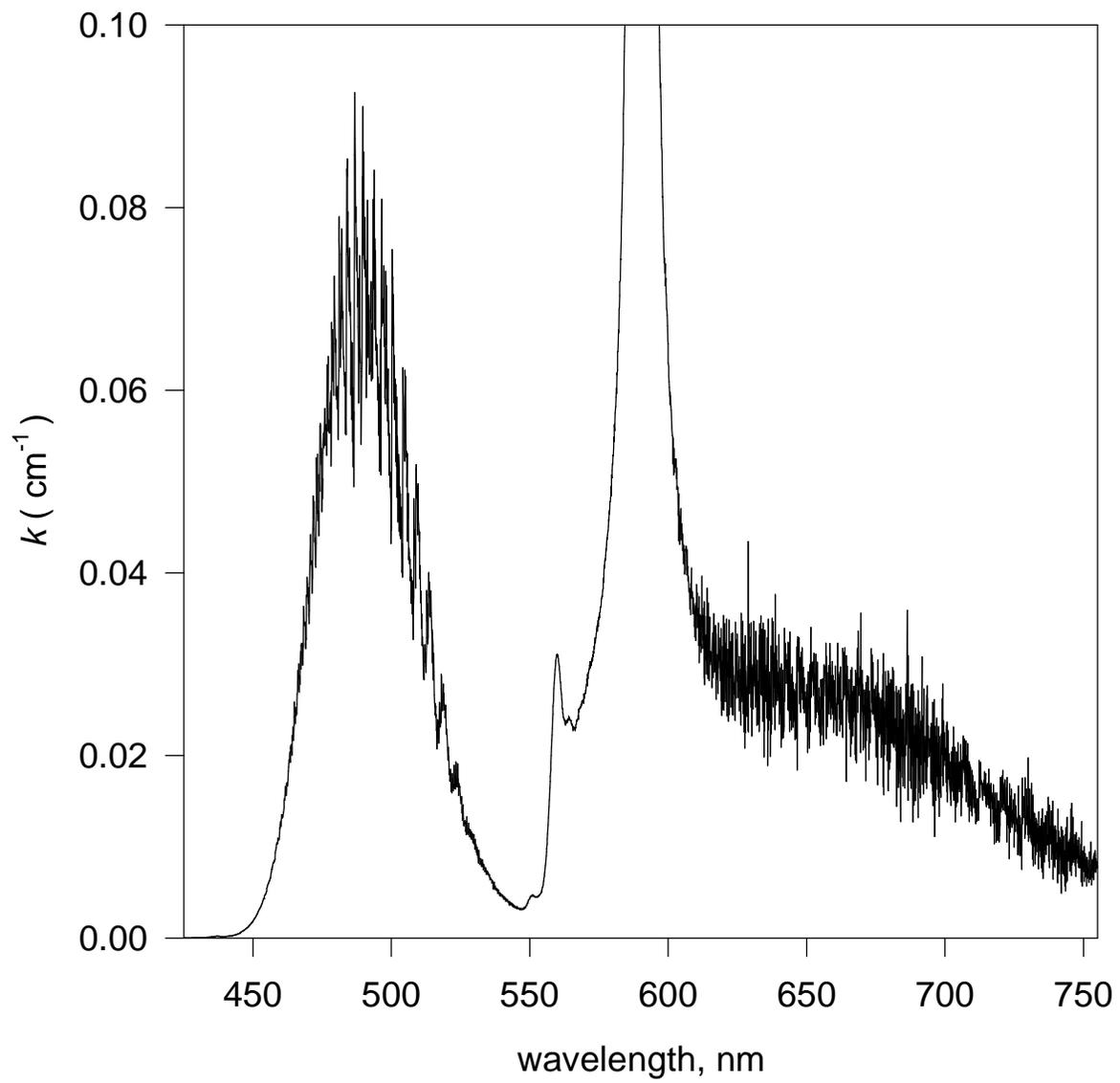
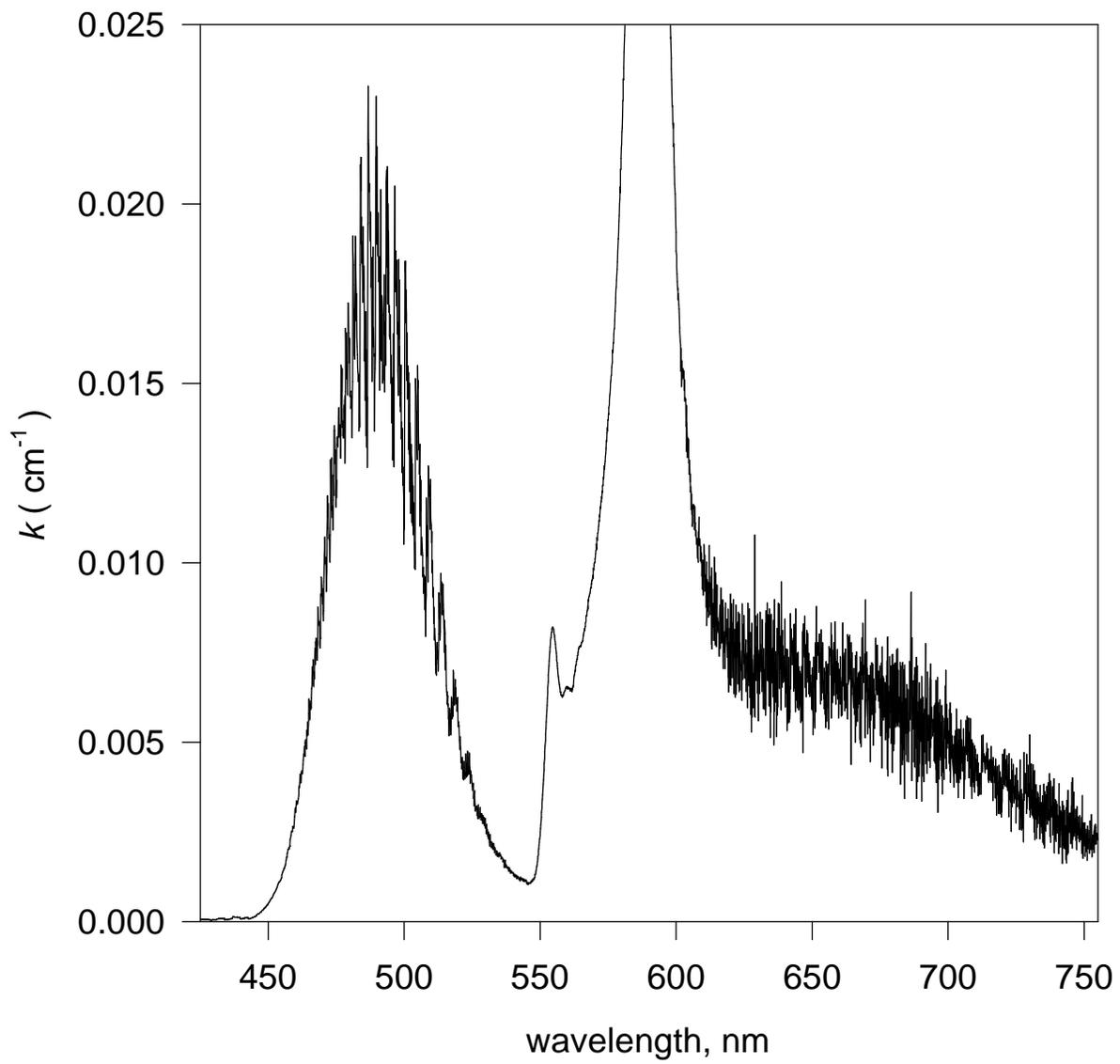
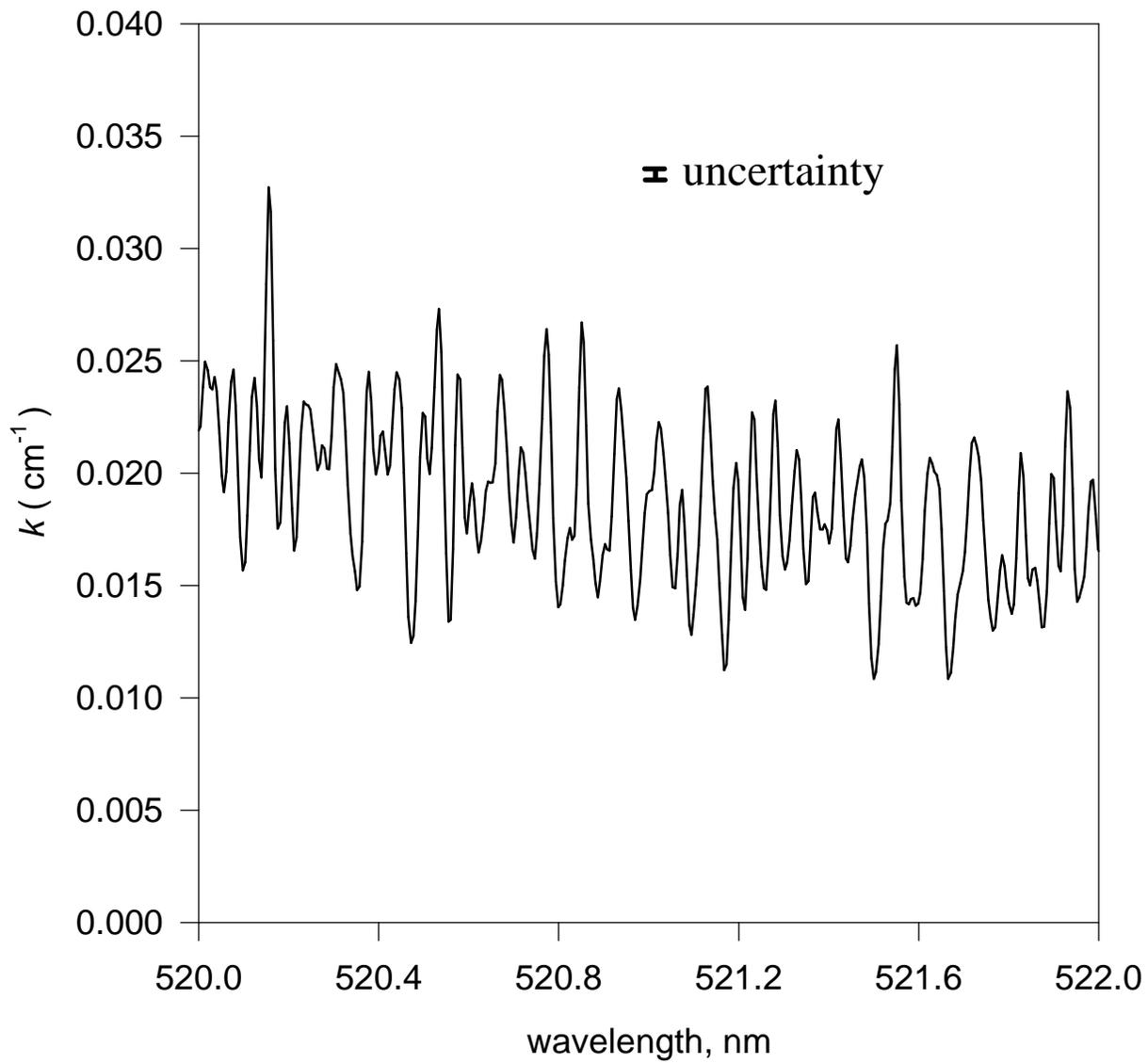
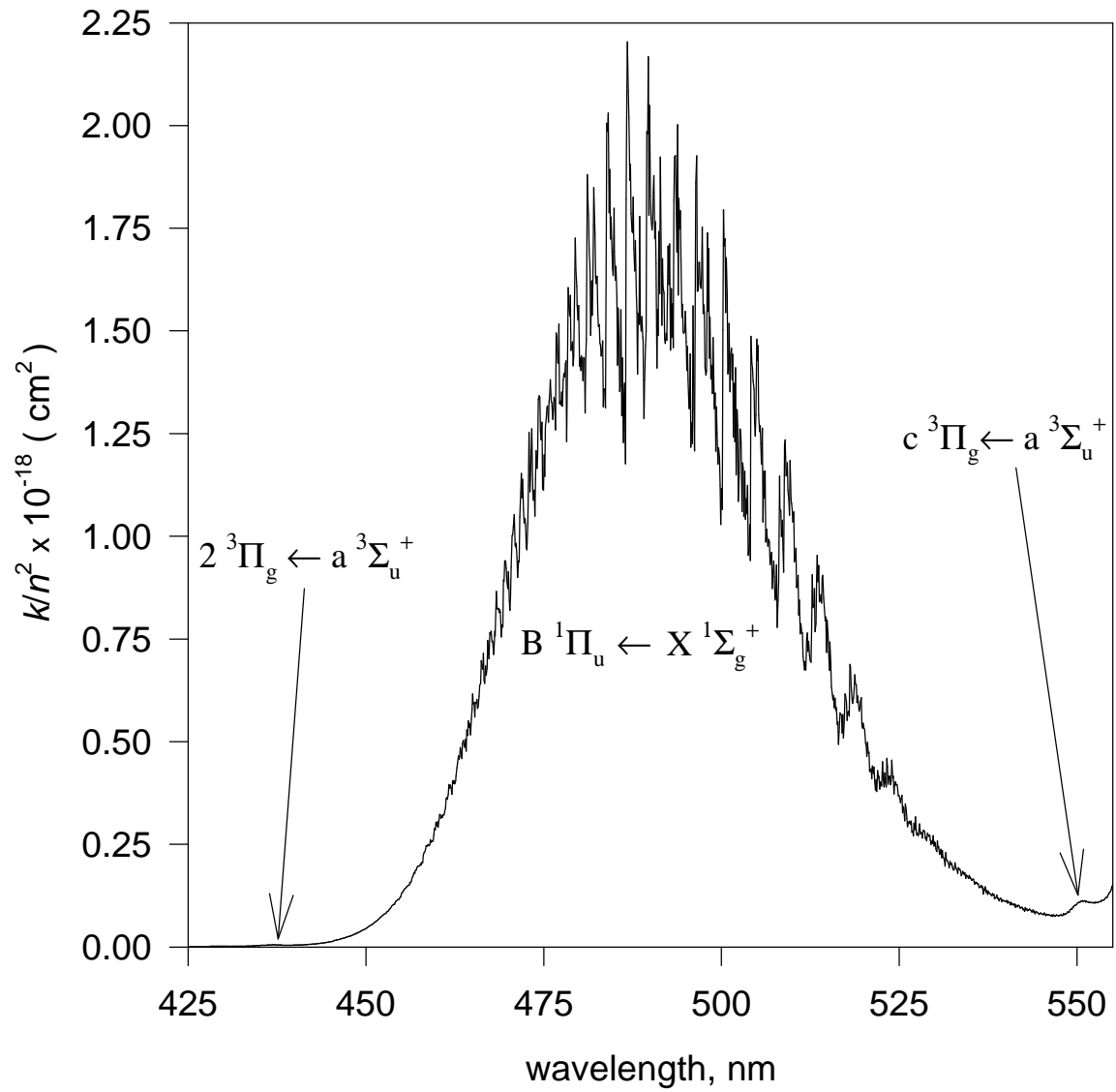
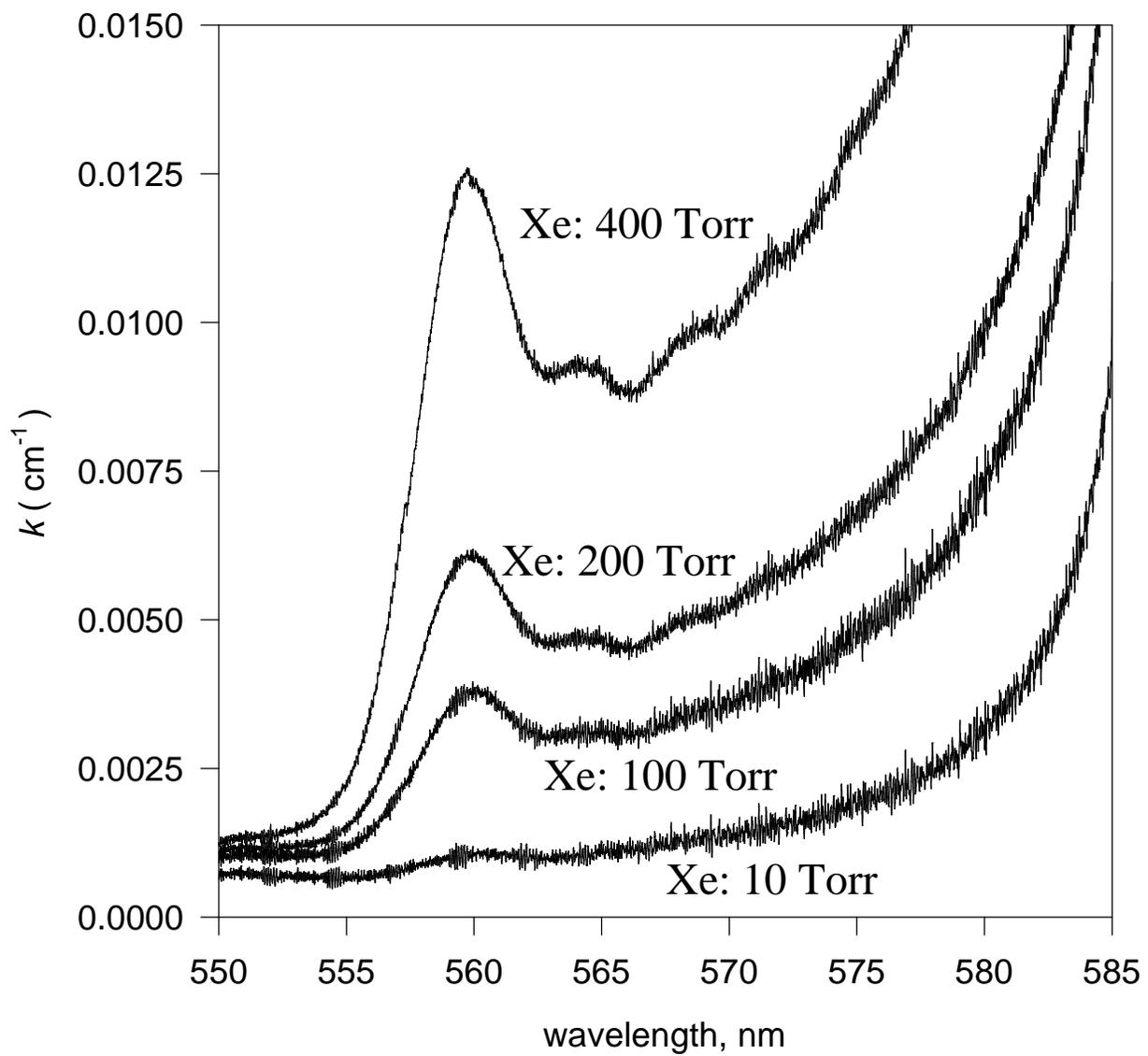
\end{document}